\documentclass[modern]{aastex631}
\newcommand{\alf}{Alfv\'en}
\newcommand{\sm}{$\sim$}

\newcommand{\ha}{H$\alpha$}

\begin{document}
 
\title{Solar Alfv\'enic Pulses and Mesoscale Solar Wind}


\author[0000-0002-5865-7924]{Jeongwoo Lee} 
\affiliation{Institute for Space Weather Sciences, New Jersey Institute of Technology, Newark, NJ 07102, USA}
\affiliation{Center for Solar-Terrestrial Research, New Jersey Institute of Technology, Newark, NJ 07102, USA}
\affiliation{Big Bear Solar Observatory, New Jersey Institute of Technology, Big Bear City, CA 92314, USA}

\author{Manolis K. Georgoulis}
\affiliation{Johns Hopkins Applied Physics Laborarory, 11100 Johns Hopkins Road, Laurel, MD 20723, USA}
\affiliation{Research Center for Astronomy and Applied Mathematics, Academy of Athens, 11527 Athens, Greece (on leave)}

\author{Rahul Sharma}
\affiliation{Department of Mathematics, Physics and Electrical Engineering, Northumbria University, Ellison Pl, Newcastle Upon Tyne, NE1 8ST, UK}

\author{Nour E. Raouafi}
\affiliation{Johns Hopkins Applied Physics Laborarory, 11100 Johns Hopkins Road, Laurel, MD 20723, USA}

\author{Qin Li}
\affiliation{Institute for Space Weather Sciences, New Jersey Institute of Technology, Newark, NJ 07102, USA}
\affiliation{Center for Solar-Terrestrial Research, New Jersey Institute of Technology, Newark, NJ 07102, USA}
\affiliation{Big Bear Solar Observatory, New Jersey Institute of Technology, Big Bear City, CA 92314, USA}

\author{Haimin Wang} 
\affiliation{Institute for Space Weather Sciences, New Jersey Institute of Technology, Newark, NJ 07102, USA}
\affiliation{Center for Solar-Terrestrial Research, New Jersey Institute of Technology, Newark, NJ 07102, USA}
\affiliation{Big Bear Solar Observatory, New Jersey Institute of Technology, Big Bear City, CA 92314, USA}

\begin{abstract}
Large-scale solar ejections are well understood, but the extent to which small-scale solar features directly influence the solar wind remains an open question, primarily due to the challenges of tracing these small-scale ejections and their impact. Here, we
measure the fine-scale motions of network bright points along a coronal hole boundary in high-resolution {\ha} images from the 1.6m Goode Solar Telescope at Big Bear Solar Observatory to quantify the agitation of open flux tubes into generating {\alf}ic pulses. 
We combine the motion, magnetic flux, and activity duration of the flux tubes to estimate the energy content carried by individual {\alf}ic pulses, which is \sm10$^{25}$ erg, adequately higher than the energies \sm10$^{23}$ erg estimated for the magnetic switchbacks observed by the Parker Solar Probe (PSP).
This implies the possibility that the surface-generated {\alf}ic pulses could reach the solar wind with sufficient energy to generate switchbacks, even though some of then are expected to be reflected back in the stratified solar atmosphere.
{\alf}ic pulses further reproduce for the first time other properties of switchbacks, including the filling factor above \sm8\% at granular and supergranular scales, which correspond best to the lower end of the mesoscale structure. 
This quantitative result for solar energy output in the form of {\alf}ic pulses through magnetic funnels provides a crucial clue to the ongoing debate about the dynamic cycle of energy exchange between the Sun and the mesoscale solar wind that has been raised, but has not been adequately addressed, by PSP near-Sun observations.

\end{abstract}


\section{Introduction}\label{sec1}

The question of how small-scale solar ejections may influence the inner heliosphere has always been timely since the prediction of solar wind \citep{Parker1958}. 
The solar wind velocities measured by PSP as a function of helio-distance consist of numerous jets superimposed over a lower bound curve \citep{Raouafi2023}. The latter curve seems to indicate the base solar wind obeying the Parker's solar wind solution \citep{Parker1965}, and the former suggests small-scale impulsive ejections from solar corona.
The best visible and thus best known solar transients are coronal mass ejections (CMEs), which drive space weather \citep{2023Howard}. On the other hand, the smallest solar wind structure in the kinetic regime is probably formed en route without direct energy exchange with the Sun 
\citep{Verscharen2019}.
Between them lies the so-called mesoscale regime with scales at 1 AU in the spatial range of 5–10,000 Mm and temporal range of 10 s to several hours, which must be important for the birth of the solar wind and the base of space weather \citep{Viall2021}. The Sun's potential contribution to the mesoscale regime is the main focus of this study.

The Sun indeed has many features on the scales corresponding to the mesoscale, but it is hard to prove whether or not they can actually reach the heliosphere with sufficient energy and number, because of technical difficulties in tracking them from the Sun's atmosphere to the solar wind. This said, the Parker Solar Probe \citep[PSP;][]{Raouafi2023SSRv} flies sufficiently close to the Sun to amply detect a special feature called switchbacks, magnetic field deflections \citep{Bale2019,Kasper2019,deWit2020,Krasnoselskikh2020}. 
Switchbacks are found in the MHD regime down to the ion gyroperiod  \citep{deWit2020, Larosa2021, Viall2021, Shi2022, 2025arXiv250608278C} with the majority lying in the mesoscale \citep[Figure 6 of][]{Larosa2021}. While identifying exclusive boundaries of the switchback scales needs further work on extracting pure switchback signals \citep[e.g.,][]{deWit2020}, switchbacks are introduced as an important structure in the mesoscale \citep{Viall2021}.
The solar features that we present in this study are in the MHD scale well below the CME scale, and thus, if found appropriate for the switchbacks, may provide a clue to the solar influence upon the mesoscale solar wind. 
The defining properties of switchbacks indicative of its solar origin include the modulation of groups of switchbacks at supergranulation scales and of individual features at granulation scales \citep{Fargette2021,Bale2023,Bale2021, Leej2022}, an aspect ratio and occurrence rate similar to that of spicules \citep{Leej2022}, and a base magnetic polarity distribution near coronal hole boundaries \citep{Leej2024}. Yet unexplored quantities such as the energy and filling factor of the {\alf}ic pulses could also play a critical role, as theses quantities measured in the Sun can be compared with those in the solar wind.

When searching for solar sources of solar wind transients, coronal hole boundaries are widely considered a key origin of switchback events \citep[e.g.,][]{Bale2019, Rivera2024}, where the photospheric magnetic field expands laterally and radially, establishing a direct connection between the solar surface and the interplanetary magnetic field.
This open magnetic topology lacks significant perpendicular inhomogeneities, thereby limiting interactions between outward-propagating wave pulses and the surrounding plasma. Under these conditions, nearly incompressible waves primarily undergo longitudinal–transverse mode coupling 
\citep{Ulmschneider1991, Kalkofen1997}, which can gradually amplify wave amplitudes and lead to the development of large-amplitude {\alf}ic waves, reported as potential switchback candidates \citep{Rivera2024}. A relationship has been suggested between the observed root-mean-square velocity amplitudes of transverse {\alf}ic waves at different heights in the solar atmosphere and in the interplanetary medium \citep{Wangs1990, Cranmer2005}, by which the observed wave amplitudes can be scaled through expanding magnetic fields under the pressure balance between a magnetic flux tube and the surrounding atmosphere. The spatial scales of coherent {\alf}ic waves observed in the corona \citep{Sharma2023} and those of switchback patches \citep{Bale2019} are also comparable to supergranulation scales that could further be scaled down to measured wave energy injection scales in the photosphere \citep{Abramenko2013} for uniformly expanding magnetic fields. This correspondence provides compelling evidence for a strong connection between the in-situ characteristics of switchbacks and their origins in the solar photosphere.

Recent PSP observations have further elucidated the processes of {\alf} waves as solar origin for switchbacks, including the distinction between solar surface-originating and in situ-generated mechanisms for {\alf}ic switchback waves  \citep{Rivera2024}, tracing switchbacks to the Sun using PSP’s perihelion
encounters \citep{Bowen2025, Bizien2025, Larosa2024} as well as simulating the generation of {\alf} waves in the Sun and their propagation into the solar wind  \citep{Wyper2024AGU01W, Touresse2024}.
This increasing number of
studies on {\alf} waves for switchbacks raise another timely question: under what conditions are {\alf}ic pulses generated in the solar surface and 
are they strong enough to lead to switchbacks?
In this Letter we study wave generation by directly measuring the fine-scale motion of the open field footpoints in the photosphere, under the premise --in lack of a definitive answer--
that photospheric motions  agitate field lines to excite {\alf} waves and that waves traveling along the open fields into space can possibly influence the mesoscale solar wind. Unlike previous studies investigating the wave origin of switchbacks \citep{Squire2020, Shoda2021,  Mallet2021, Squire2022, Johnston2022}, we focus on the solar conditions required to generate {\alf}ic waves with sufficient energy to produce the large deflections observed by PSP.

\begin{figure}[tbh]  
\plotone{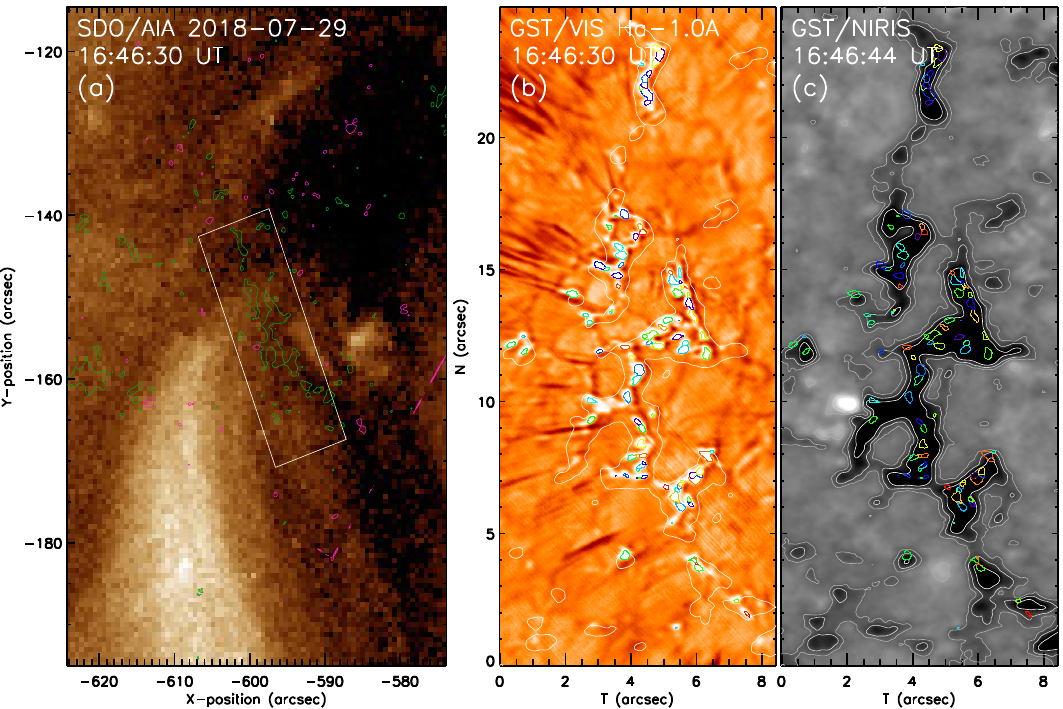}
\caption{Target region in the coronal hole boundary. 
(a) SDO/AIA 193 {\AA} image shows the coronal hole and EUV bright regions around. The superimposed contours represent the LoS magnetic field at +50 G (pink) and $-$50 G (green) from the NIRIS magnetogram. The white box denotes the FOV of the other panels. 
(b) H$\alpha$ far blue wing image in the sub-region shows spicules (dark straw-like features),  filigree (white contours), and NBPs (crinkle-like features outlined by the colored contours). This field of view has the ordinate in the solar radius direction (named `$N$') and the abscissa in the azimuthal direction (`$T$'). 
(c) NIRIS LoS magnetogram as the gray background image and contours at  $[-100, -50, -25, 25, 50]$ G. The NBPs (colored contours) appear only in single polarity fields.
An animation (f1.mp4) is available showing this figure for the entire period of investigation from 16:31:40 UT to 18:06:59 UT.}
\label{fig:1}
\end{figure}

\section{Data}\label{sec3}

Essential in this study are high-resolution images and a powerful feature tracking tool. 
We use the high-resolution {\ha} images obtained from the Visible Imaging Spectrometer \citep[VIS;][]{Cao2010} and magnetograms from Near-infrared Imaging Spectropolarimeter \citep[NIRIS;][]{Cao2012} mounted on the 1.6 m Goode Solar Telescope \citep[GST;][]{Goode2012} at Big Bear Solar Observatory (BBSO).
The data were obtained on 2018 July 29 for a quiet-Sun coronal hole boundary region located about a half solar radius from the disk center at (604$''$ E, 125$''$ S). 
During this observation between 16:34–18:38 UT, NIRIS took high spatial resolution (0.24$''$) magnetograms and the VIS provided high resolution (0.10$''$) {\ha} images at 11 wavelength points between $\pm$1{\AA} in interval of 0.2 {\AA} in the {\ha} line center. 
The time interval between consecutive sets of the 11 wavelength point observation is 40 s. This cadence is adequate for the present analysis, because previous studies have reported the periodicity of transverse oscillations associated with MBPs in the range of 130--400 s \citep{Jess2012, Berberyan2024}.
From the coaligned NIRIS magnetograms one can then determine the line-of-sight (LoS) magnetic field or de-project it for an estimate of the vertical magnetic field, $B_z$ (see also Figure \ref{f4}). 

Figure \ref{fig:1} shows a region of the chromospheric network near the coronal hole boundary, as can be identified with the SDO/AIA 193 {\AA} image (Figure \ref{fig:1}a). The fine structure of the network boundary is visible in the high resolution {\ha} far wing image of GST/VIS (Figure \ref{fig:1}b). The straw-like dark features are spicules \citep{Secchi1877} and the bright lanes underneath spicules are filigree \citep{DunnZir1973}. The filigree consisted of many crinkles called Network Bright Points (NBPs), also called Magnetic Bright Points (MBPs). The brightness is generally believed to result from a reduction in opacity when plasma escapes along open field lines \citep{DunnZir1973}.
The NIRIS magnetogram is shown in Figure \ref{fig:1}c as a background grayscale image, and the overplotted colored polygons representing NBPs are taken from Figure \ref{fig:1}b. It is notable that these NBPs lie within the unipolar magnetic fields concentrated along the coronal hole boundary  (Figure \ref{fig:1}c).
NBPs move in network fields due to an evolving magneto-convection pattern \citep{Ballegoo1998, Nisenson2003, Utz2010, Chitta2012}, and this motion is believed to excite flux tube waves, which propagate up to the corona and even the solar wind \citep{Cranmer2005, Cranmer2019, Soler2019}.

\section{Results}\label{sec2}

\subsection{Dynamics of NBPs}

\begin{figure}[tbh]  
\plotone{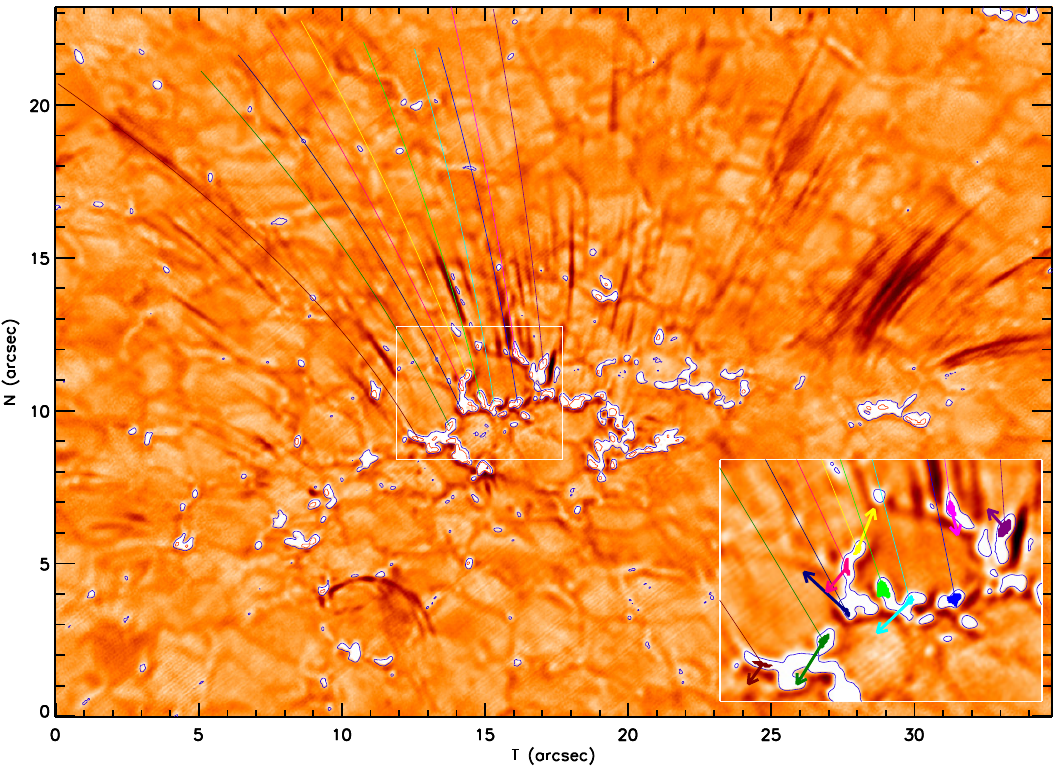}
\caption{Spicules and NBPs in a GST/VIS {\ha}-1.0{\AA} blue wing image. A few field lines inferred from the selective spicule trajectories are marked in color together with NBPs outlined by a blue contour underneath.
The inset magnifies the white box region, in which the color-filled masks represent the NBPs and associated field lines are colored in the same code. Using SWAMIS, we track each NBP in subsequent frames to determine its velocity (arrows in the inset).}
\label{fig:2}
\end{figure}

Figure \ref{fig:2} shows spicules and NBPs over a {\ha} blue wing image. Hypothetical field lines are set on the basis of the spicule trajectory and plotted to illustrate the open field lines. 
The inset shows the selected NBPs underneath the field lines. Their random motions, as evident in the high-resolution {\ha} movie, may agitate those field lines to generate {\alf} waves.
In the solar atmosphere, {\alf} speed increases with height and the waves generated with a small amplitude in the photosphere will rapidly be amplified with height. 
Due to gradients in both density and magnetic field strength in the coronal hole boundary, {\alf}ic fluctuations steepen, which may ultimately give rise to
switchbacks \citep{Shoda2021}. The open field lines in this figure as a whole will then be able to reproduce a group of switchbacks on the size of a supergranule with the individual size of a granule \citep{Fargette2021, Bale2021}.

An important question here is not whether this motion can generate {\alf} waves, but whether these {\alf} waves carry the energy necessary to be detected by PSP if they manage to reach the PSP location.
To answer this, a quantitative measurement of the subtle motions of NBPs is needed. 
We, however, found this 
challenging because many NBPs look alike and are closely seated. We use the Southwest Automatic Magnetic Identification Suite \citep[SWAMIS;][]{DeForest2007} for detection of NBPs in the VIS data, although it is an algorithm originally developed for tracking variable magnetic features near noise levels. SWAMIS first segments the filigrees and stores the results in ``masks'' such as the color-filled contours shown in the inset. 
Each mask is traced through subsequent frames, by which we can determine the motion and area of NBPs from their birth to death, and accordingly their lifetimes.  
Typical lifetimes of NBPs range from 40 s to a few min, close to the period of spicules \citep{OkaDePon2011}, and therefore the waves generated are like pulses \citep{Dover2021, Dover2022}.  
They are generated by the transverse displacement of the flux tubes and can be identified as $m=1$ kink mode waves, while {\alf} waves refer to a pure $m=0$ mode with a strong torsional component \citep{Dover2022}. 
We hereafter call them {\alf}ic pulses instead of {\alf} waves.

\begin{figure}[tbh]  
\plotone{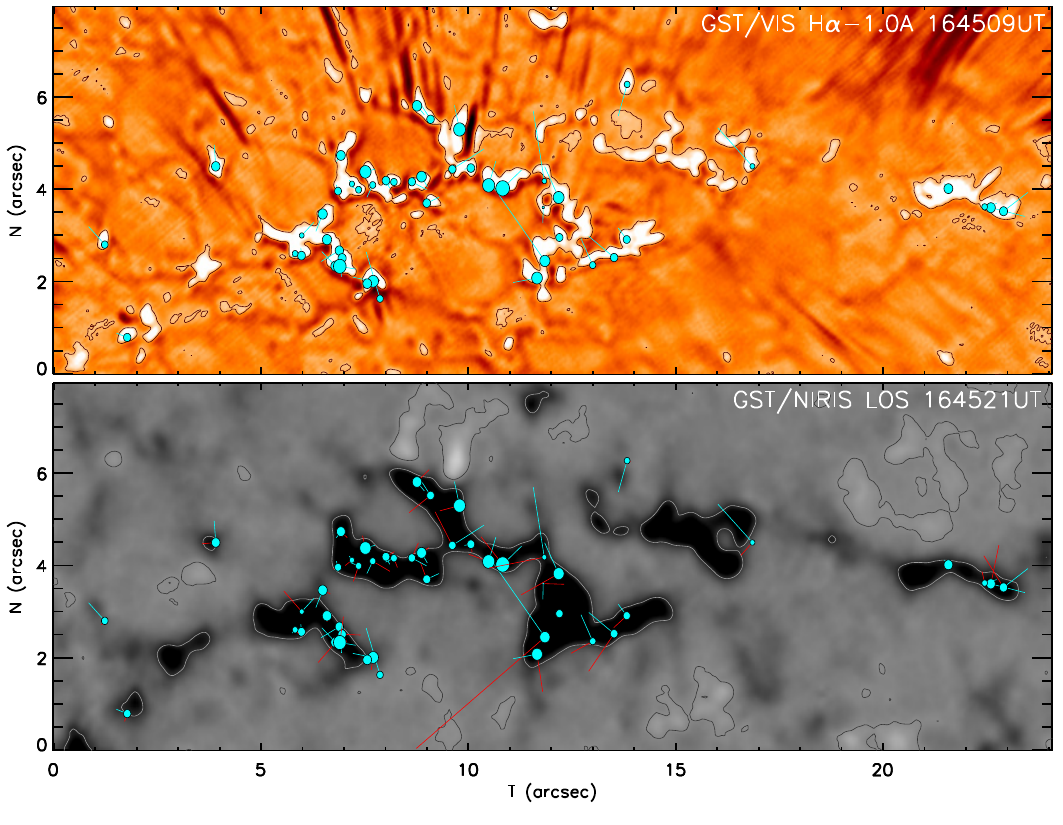}
\caption{Tracking NBP motions in a magnetogram. 
In the top panel, NBPs are marked with cyan circles with equivalent area and their total displacements during their lifetimes are indicated by cyan lines (same in both plots), with length 3 times longer than actual displacements for visual convenience. 
The equivalent areas (cyan dots) and velocities (cyan lines) of these NBPs are further shown against the corresponding NIRIS LOS magnetogram, used to calculate the electric field vectors (red lines) shown in the bottom image. 
}
\label{fig:6}
\end{figure}

\subsection{Feature Tracking}
The top panel of Figure \ref{fig:6} shows NBP areas and motions determined using SWAMIS from the GST/VIS {\ha}$-$1.0 {\AA} images. 
The feature recognition algorithm requires discrimination to separate the foreground features from the background noise. We define NBPs with the threshold for
the {\ha} intensity contrast, $(I-I_b)/I_b \geq$ 0.07.
As a result, NBPs have irregular shaped areas, which SWAMIS saves as masks (Figures \ref{fig:1} and \ref{fig:6}). We represent each NBP by a circle at its center-of-mass position ${\bf x}^k$ with an area $A^k$ of the mask. 
By tracing  ${\bf x}^k$  of an NBP with the same ID number, $k$, in consecutive frames, we measure the displacement of the NBP and convert it to  velocity using ${\bf u}^k = ({\bf x}^{k+1}-{\bf x}^k)/\Delta t$. Typical displacements of the NBPs  from frame to frame ($\Delta t=$40 s) are rather small, and we multiplied ${\bf u}^k$ by a factor of 3 for better visibility. 
(${\bf u}^k,~{\bf x}^k$) from the top panel are copied to the second panel, the GST/NIRIS magnetogram.
Upon co-aligning the masked VIS pixels over simultaneous NIRIS magnetograms, two quantities are immediately available: the magnetic flux (cyan dots) of the $k-$th NBP, $\phi_z^k= B_z^k A^k$ and the convective electric field $E_\bot^k = u^kB_z^k/c$  (red line) created by its motion (cyan line). The latter amounts to the transverse amplitude of the generated {\alf}ic pulse.
Note that we are measuring only the rectilinear motions of the NBPs to calculate the {\alf}ic pulse energy. NBPs may show additional variations such as expansion, shrinkage, or deformation of their morphology. These motions may generate torsional {\alf} waves \citep{Sharma2023, Wedemeyer2012}, which we are unable to measure with the present data and technique.

\subsection{{\alf}ic Pulse Energy and Switchback Energy}

\begin{figure}[tbh]  
\plotone{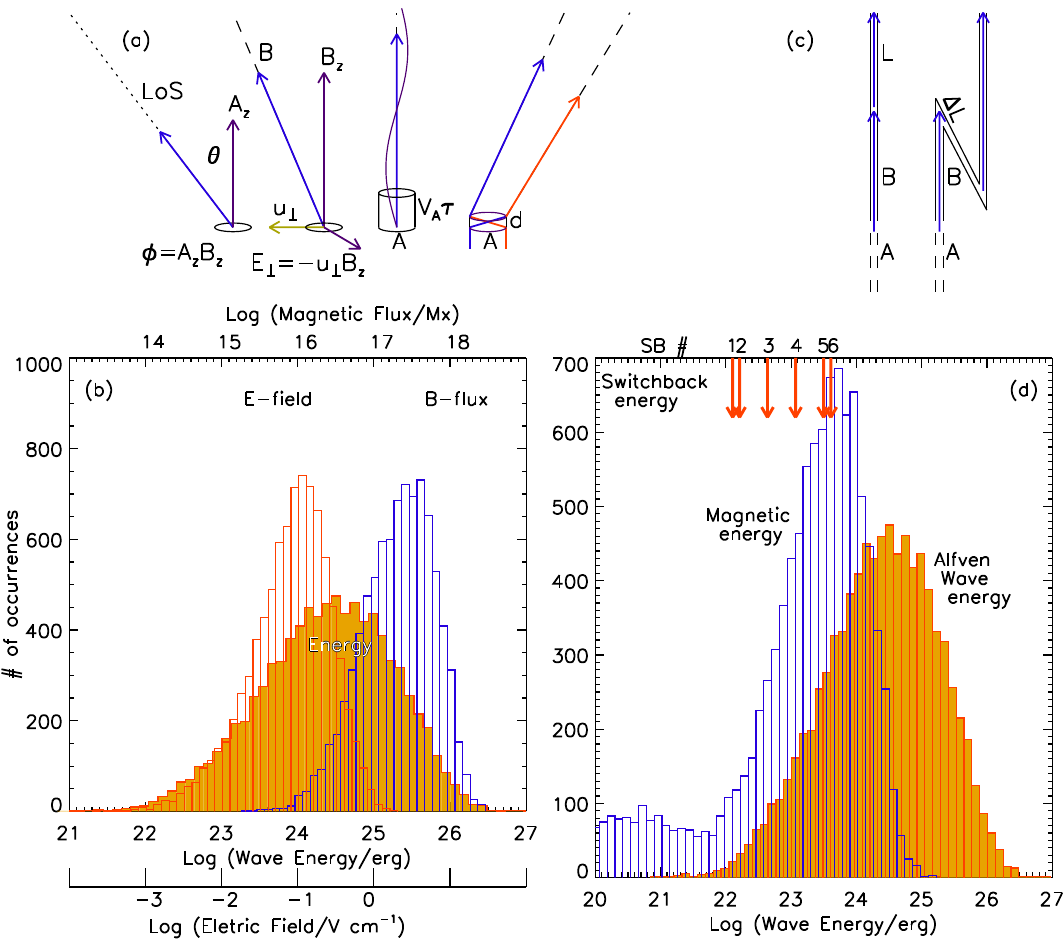}
\caption{Illustration of the energy calculation and the associated parameters (top) and the distribution of the results. (a) Parameters in the photophere and chromosphere. (b) Parameters of a magnetic switchback.  (c) Number distributions of electric field, magnetic flux, and energy of {\alf}ic pulses. (d) Energies of {\alf}ic pulses (yellow) and those associated with magnetic reconnection (blue) in comparison with six switchback energies (red arrows). SB\#1--6 refer to the PSP switchbacks with ID: E2S4, E2S6, E1S5, E1S6, E1S1, and	E2S3  defined in \cite{Laker2021}. }
\label{f4}
\end{figure}

We calculate the {\alf}ic pulse energy on an individual basis by integrating the {\alf} wave flux over area and time:
\begin{equation}
{\cal E}_A\equiv \int\!\!\int F~da~dt  
\quad {\rm with} \quad
F = \frac{1}{2}\rho u^2 V_A
\end{equation}
For this energy to
be useful for comparison with that of switchbacks under the wave origin hypothesis, 
the area integral $\int da$ should be over each fluxtube cross section, and time integral $\int dt$ over the lifetime of each driver, i.e., mobile NBPs. 
Figure \ref{f4}a illustrates how we obtain these quantities from observations. The dotted line is the LoS, and the dashed lines are magnetic field lines that connect from the solar surface (a) to the solar wind (c).
The angle, $\theta$, represents the inclination of the local normal vector with respect to the LoS, and is $\approx 39^{\circ}$ from the disk location of this FOV. We divide the observed LoS component of the magnetic field by $\cos \theta$ to obtain local $B_z$, which is then used to calculate the two major quantities: magnetic flux of individual NBPs, $\phi_z= B_z A_z$, and the convective electric field $E_\bot = uB_z/c$ created by their motion, as displayed in Figure \ref{fig:6}. The inclination of a field line is not used in this calculation, because we are using quantities perpendicular or parallel to the normal vector to the cross sectional area (denoted as $A_z$) of each flux tube lying on the photospheric surface. The NBP motion $u_\bot$ is also confined to the surface and $\cos \theta$ is used to correct the projection effect.

With $\phi_z$ and $E_\bot$ already calculated, we are left with the mass density $\rho$, the only quantity not directly obtained from this observation. For this purpose, we assume equipartition between the magnetic energy density and the ambient turbulent energy density,    $\rho u^2=B_z^2/4\pi$,
so that the energy of an {\alf}ic pulse is calculated by 
 ${\cal E}  = c\tau E_\perp \phi_z/{8\pi}$, where $\tau$ is the duration of NBPs.
We emphasize here that energy equipartition is regarded as suitable for active regions, i.e., kG fields 
\citep{2012Georgoulis}. 
However, the NBPs in the current study correlate with fields of a few hundred Gauss 
and are more mobile, so the ambient turbulence may be stronger than the magnetic field. 
In such environments, one expects a plasma $\beta > 1$, so assuming equipartion implies an underestimation of the density $\rho$ and hence of the energy of \alf ic pulses. As it stands, our analysis provides a lower limit of $\cal E$ which is physically meaningful in our context.

It is desirable to calculate the amount of energy associated with magnetic reconnection as well. This requires additional information not only on NBPs but on their opposite polarity pairs and magnetic configuration. Although a more detailed study on this topic is underway \citep{Manolis25}, we adopt a simplified assumption that all flux tubes attached to NBPs reconnect with their pairs and completely annihilate over a cylindrical volume with each side equal to the radius of the NBP. In this case, the
magnetic energy to be released from an annihilation volume, $\pi d^3$, is
\begin{equation}
{\cal E}_B\equiv  \frac{B^2}{8} d^3.
\end{equation} 
In reality, neither all NBPs can reconnect with each other nor magnetic field inside an NBP would completely annihilate. 
This crude calculation is intended to give an upper bound for the magnetic energy available in the photosphere.
Note that we use the same magnetic field and the fluxtube cross section, $\pi d^2$, for both types of energy, and we should clarify what makes them different.  As illustrated in Figure \ref{fig:6}a, the magnetic energy is calculated by assuming the annihilation length of $d$. For the wave energy calculation, $d$ is practically replaced by $\tau V_A$. The propagation speed $V_A$ is determined by the medium and $\tau$ by the driver so that $\tau V_A$ may differ from $d$, and so does ${\cal E}_A$ from ${\cal E}_B$.

Figure \ref{f4}b shows the result of the above calculation.
Since not all NBPs are alike, we plot their individual energies (orange area) as well as fluxes (blue bars) and electric fields (red bars) as number distributions.
The magnetic flux of individual NBPs is distributed around a few 10$^{17}$ Mx, one order of magnitude lower than the small-scale magnetic fluxes in the quiet Sun \citep{Parnell2002,Parnell2009}. This makes sense because NBPs occupy a small fraction inside the network fields. 
The convective electric field $E_\bot$ is around $0.1-1.0$ V cm$^{-1}$, which is is about a hundred times smaller than the typical electric field associated with solar flares as estimated from flare ribbons \citep{2022LeeRvMPP}.
Both the magnetic flux and the electric field are narrowly distributed, but they together make the wave energy in a broader distribution in the range of $10^{23} - 10^{26}$ erg with peak frequency of occurrence at $7\times 10^{24}$ erg. 

To our knowledge this is the first time energies of  individual {\alf}ic wave pulse energies are calculated, whereas other studies typically calculate the energy flux density of {\alf}ic pulses. Our estimate for the energy flux is around 5 kW m$^{-2}$ higher than reported elsewhere, for instance, 250--440 W m$^{-2}$ \citep{OkaDePon2011, Wedemeyer2012}.

For the first time, we calculate the switchback energy using the simple scheme illustrated in Figure \ref{f4}c. The N-shaped switchback is a schematic representation of a highly steepened {\alf}ic pulse.
The energy used to bend the fluxtube should be proportional to the increased length, $\Delta L$. If the fluxtube has field strength $B$ and area $A$, this energy is then expressed as $\Delta {\cal E}= B^2AL/8\pi = 
\alpha B^2D^3/32$ where $D$ is diameter and $\alpha=L/D$ is the aspect ratio.
We obtain these parameters for the six switchbacks listed in Table A.1 of \cite{Laker2021} and use them to find the magnetic energies in the range of $5\times10^{21} - 3\times10^{23}$ erg. The kinetic energy should be added to get the total energy of the switchbacks so that those
plotted as red arrows in Figure \ref{f4}d include a multiplicative factor of two.
They are lower than the {\alf}ic pulse's energy  $10^{23} - 10^{26}$ erg
so that we can entertain a scenario that {\alf}ic pulses impart their energy to switchbacks
\citep{Wedemeyer2012}.
The magnetic energies are also shown as blue bars in Figure \ref{f4}d, which lie in the range of $10^{19} - 10^{25}$ erg to overlap with the switchback energies (red arrows). If not all NBPs reconnect as we assume, then the actual number in each energy bin should reduce, albeit not the range of the energy distribution. If there were more numerous opposite-polarity patches in undetectably small sizes, the number in each energy bin would increase, but the dominant energy range can fall below that of switchbacks. 

An unavoidable factor in the above energy comparison is the possible reflection of {\alf}ic pulses in the solar atmosphere due to stratification of the density and magnetic field. On one hand, it makes the small amplitudes of the {\alf}ic pulses in the photosphere rapidly increase with height, which can also lead to steepening of {\alf}ic fluctuations, which could then create kinked structures such as switchbacks.
On the other hand, the corresponding change in {\alf} speed can cause reflection of {\alf}ic pulses, especially at the transition layer
\citep{Hollweg1978,Schwartz1984,Cranmer2005,ChaeLee2023}. 
We calculated the reflection coefficient in Appendix \ref{ap:1} to find that  1--10\% of the {\alf}ic pulse energy can be transmitted through the transition region. 
This means that 
the range of transmitted energy will then shift down by one or two octaves with their number maintained. This energy distribution still overlaps with the energies of the switchbacks.

\begin{figure}[tbh]  
\plotone{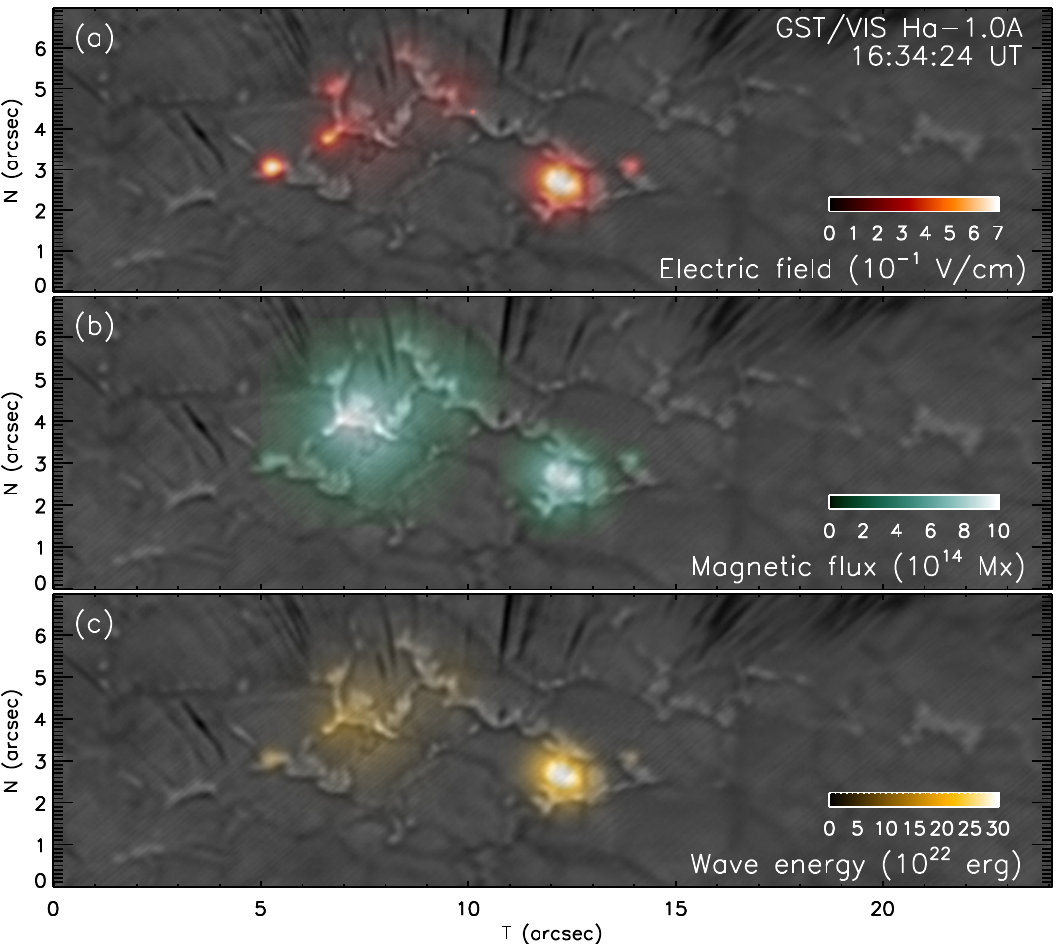}
\caption{Physical quantities derived from the GST/VIS {\ha} blue wing images and the GST/NIRIS
magnetograms. (a) Convective electric field, (b) magnetic flux of NBPs, and (c) energy of {\alf}ic pulses are plotted as color shades of Gaussians centered on individual NBPs.
An animation (f5.mp4) of this figure is available for the period from 16:31:42 UT to 18:06:03 UT with the logarithmic scale bars instead of the linear scale bars as shown here.
}
\label{f3}
\end{figure}

\subsection{Spatial Distribution and Filling Factor of {\alf}ic Pulses}

Although the magnitudes of the flux $\phi_z$, convective electric field $E_\bot$ and energy ${\cal E}$ of {\alf}ic pulses
are the ultimate goal of this study, their discrete spatial distribution and subsequent filling factor are also of interest for comparison with switchbacks measured in the form of time series. To visualize the spatial distribution of these quantities, we represent them as Gaussians centered at the position of each individual NBP, wide in proportion to the diameter of the NBP and high in proportion to the magnitude of the quantities and plot them as colored shades over the {\ha} far-wing image in Figure \ref{f3}. 
The Gaussian assumption is only for visualization and does not affect the calculation made in the previous section.
The magnetic fluxes of the NBPs (Figure \ref{f3}b) and the energies of individual pulses (Figure \ref{f3}c) are plotted in the same format. Comparison of the three panels in Figure \ref{f3} reveals that regions with the strongest electric field or magnetic flux are not necessarily the sites of the highest energy because mobile NBPs and strong field regions do not necessarily move in tandem. 
In general, NBPs moving in strong field regions produce intense electric field, while those in weak field region do not. Accordingly strong wave sources are more limited than NBPs themselves.
In the accompanying animation, each frame shows a scale bar for the magnitude range of the plotted quantity, meant to show its variation in time.

\begin{figure}[tbh]  
\plotone{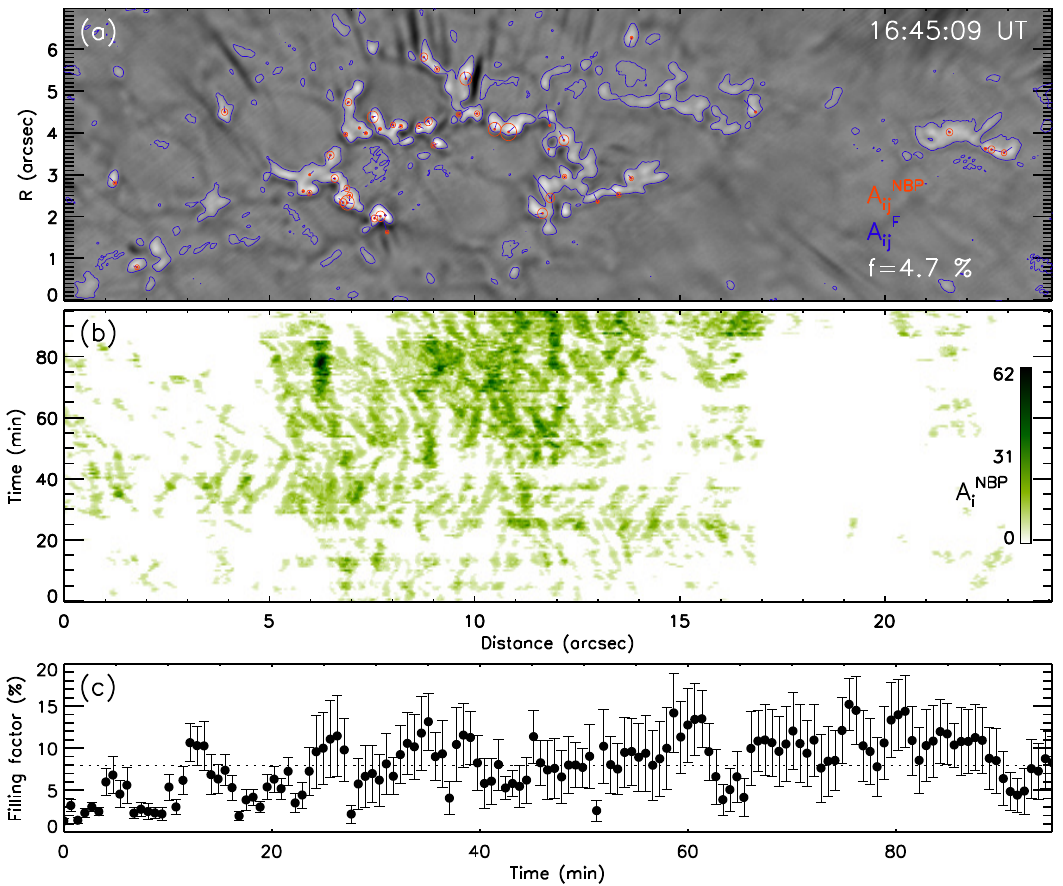}
\caption{Filling factor of NBPs. (a) The areas of mobile NBPs (red circles) and filigree (blue contours)  The filling factor is calculated by dividing the sum of the NBP areas by the total area of the filigree at each time. (b) Time-distance (T-R) map of the NBPs. At each time, the NBP areas along the $T$-axis is obtained by adding up the NBPs along the $R$-axis. 
(c) Average filling factor as a function of time. The error bars are based on the uncertainty in the filigree area and the dashed line at $\sim$8\% is the mean value of the filling factor over time. 
}
\label{f5}
\end{figure}


The filling factor of NBPs, namely, how densely NBPs cover the solar surface, is another important constraint in the current {\alf} wave/turbulence models \citep{Squire2020, Squire2022, Mallet2021, Shoda2021, Johnston2022}. 
PSP observations show a filling factor as high as 6\% \citep{Bale2019}, which is not well understood.
We can reevaluate this parameter more carefully using the SWAMIS result.
In Figure \ref{f5}a, we mark the areas of mobile NBPs with red circles with equivalent areas in which by $A_{ij}^{NBP}=1$ and zero otherwise, where $i$ and $j$ denote the pixel indices along the $x$- and $y$-axes, respectively. All filigree structures—including both mobile and immobile NBPs—are outlined with blue contours, inside which $A_{ij}^F=1$ and zero outside. The latter represents the open field region which forms the solar wind with and without switchbacks. 
Under the current hypothesis, the field lines stemming from pixels with $A^{NBP}_{ij}=1$ develop into switchbacks, and those from $A^{F}_{ij}=1$ remain as straight field lines forming quiescent non-switchback regions in the solar wind. Therefore, the ratio of the total area of the mobile NBPs to that of the filigree:
\begin{equation}
    f \equiv \frac{\sum_{i,j} A_{ij}^{NBP}}{\sum_{i,j} A_{ij}^F} .
\end{equation}
should give the filling factor. 

A single filling factor is obtained per image, with the positional information integrated out. To show the spatio-temporal distribution of NBPs, we calculate $ A_{i}^{NBP} = \sum_j A_{ij}^{NBP}$ on each frame and stack them along the y-axis. Such 1-D structure along the $T$-axis is calculated over all 142 frames and stacked together to build the time-distance map in Figure \ref{f5}b with the color set to be darker for larger values.
The average filling factor is finally calculated by performing the summation over the $T$-axis as well and shown in Figure \ref{f5}c as a function of time. The vertical error bars are based on the 2--3\% intensity contrast threshold for defining the filigree region. The time-averaged filling factor is now as high as 8\%, comparable to 6\% of the switchbacks detected by PSP \citep{Bale2019}. 
Our high filling factor is obtained by excluding the regions considered irrelevant for comparison with switchbacks; they are either the field-free regions or the closed-field regions.
The low filling factor predicted by {\alf} wave/turbulence models for switchbacks \citep{Shoda2021} could also arise in this case if such solar surface conditions were not accounted for.

\section{Discussion}\label{sec:3}

Using high-resolution {\ha} images and magnetograms from BBSO/GST, we have investigated whether the Sun can generate {\alf}ic pulses \citep{Asgari2013} strong enough to contribute to the meso-scale solar wind.
Our results support the {\alf} wave/turbulence models for switchbacks \citep{Squire2020, Squire2022, Mallet2021, Shoda2021, Johnston2022},
but draw more attention to the solar source properties. 
The investigated solar wind source is located in the edge of a coronal hole, where steep gradients in both density and magnetic field can result in the steepening of {\alf}ic fluctuations \citep{Wangs1990} and thus shock-like structures characterized by high velocities and abrupt changes in magnetic field strength
\citep{Dover2022, Dover2021}, favorable for the switchback formation.
In these regions, the open magnetic fields are mainly unipolar, so that {\alf}ic deflections propagating along them may be detected as a series of switchback-like structures by a flying spacecraft \citep{Leej2022, Leej2024}.
Counting the discrete NBPs yields another important clue, the filling factor of {\alf}ic pulses \sm8\% as high as detected by PSP measurements for the switchback patches \citep{Bale2019}.
These solar properties, not fully adopted by in situ only modeling \cite{Shoda2021}, augment the {\alf} wave/turbulence models for switchbacks \citep{Squire2020, Squire2022, Mallet2021, Shoda2021, Johnston2022}. 


As a crucial step, we evaluated the potential solar candidates for switchbacks on an individual energy basis.
Collection of tiny bipolar flux cancellations and the eruption of minifilaments may also contribute to the solar wind, but in smaller and larger individual energies, respectively.
The energy of individual {\alf}ic pulses lies in the range of $10^{24} - 10^{25}$ erg at generation (\S 2.3), but only a fraction of it will reach the solar wind due to strong reflection at the solar transition region. We estimated the transmission coefficient as 1--10\% using parameters (\S 2.4), and expect the energy content of the {\alf}ic pulses in space is  yet comparable to the energy range of switchbacks, $10^{22} - 10^{23}$ erg. 
The individual {\alf}ic pulse energy and the overall filling factor (\S 2.5) serve as major quantities in connecting the Sun to switchbacks.
While this study favors {\alf}ic pulses as the solar source of the switchbacks, magnetic reconnection is also a strong candidate, if not primarily for switchbacks, but as a significant contributor to the bulk energy flux for powering the solar wind. 

Since flux tubes kinked as sharply as switchbacks are rarely found in the Sun, the switchback morphology must form en route \citep{Akhavan2024} and the Sun provides only seeds for switchbacks. In this case, we must consider how the Sun generates the seeds at the appropriate scale and rate for in situ switchback formation.
The Sun as a star has the granular and supergranular structures that form under superadiabacity in sizes determined by partial ionization of hydrogen and helium \citep{Stix2004}.
As flux tubes congregate toward the convective cell boundary and are enhanced there playing a role as generators and also conduits for the {\alf} waves \citep{Simon1964, Cranmer2005}, it is an inevitable consequence that modulations in those two convective scales are found in switchbacks \citep{Bale2021, Fargette2021}. 
It is then solar convection that underlies solar input to the meso-scale solar wind.
An emerging picture is that the Sun outputs a granule-scale excess energy into the space utilizing {\alf}icity rather than direct mass transfer and the solar wind stores those small-scale excess energies in the form of magnetic switchbacks \citep{Leej2024}.

Switchbacks themselves seem to be able to heat and accelerate the solar wind nonadiabatically \citep{Rivera2024}, and further create disturbances, e.g., whislter waves, which can scatter strahl particles to regulate heat flux from the Sun \citep{Cattell2022, Choi2024}. This forms an important cycle of energy exchange between the Sun and the mesoscale solar wind.
Together with the large-scale CMEs from strong fields accumulated in active regions \citep{Zirin1988}, short-period {\alf}ic pulses generated at the network boundary should complement  the wide range of lower corona inputs to the solar wind.

\begin{acknowledgments}
We thank Drs Kyungeun Choi, Thierry Dudok de Wit, and Andrea Larosa for helpful discussions.
This work was supported by NSF grants, AGS-2114201, AGS-2229064 and AGS-2309939, and NASA grants, 80NSSC19K0257, 80NSSC20K0025, 80NSSC20K1282 and 80NSSC24K0258. RS is grateful for support from the UKRI Future Leader Fellowship (RiPSAWMR/T019891/1).
\end{acknowledgments}

\vspace{5mm}
\facilities{Big Bear Observatory (GST/VIS and GST/NIRIS)}

\software{IDL, SolarSoft \citep{2012ascl.soft08013F}, SWAMIS \citep{DeForest2007}}

\appendix

\section{Reflection of Alfv\'enic Pulses}\label{ap:1}

\begin{figure}[tbh]  
\plotone{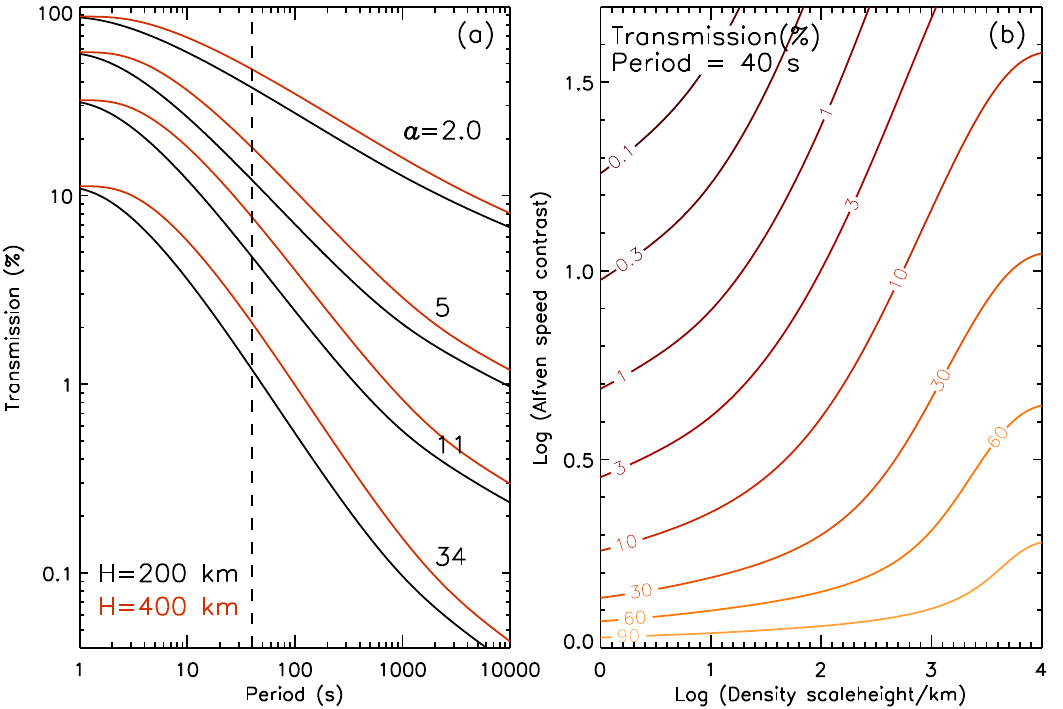}
\caption{Transmission of {\alf}ic pulses. (a) Transmission coefficient in energy as a function of period for selected values of the {\alf} speed contrast, $\alpha=V_{A_2}/V_{A_1}$ calculated for two density scale heights, $H_\rho$. (b) Transmission coefficient as a function of $H_\rho$ and $\alpha$ at a fixed period of 40 s. Different contours between
0.1\% and 90\% are labeled with these percentage values. Darker contours denote stronger wave reflection.
}
\label{f7}
\end{figure}

The amount of wave energy that can be transmitted to the corona and the solar wind depends on the gradient of {\alf} speed and the wave frequency.
{\alf}ic pulses with frequencies higher than the critical frequency determined by the gradient of the {\alf} speed, $\omega_c =|\partial V_A/\partial z |/2$, can propagate as if in a homogeneous medium,
while those with lower frequencies suffer reflection \citep{Schwartz1984}.
For a typical transition layer, $\omega_c^{-1}$ is about 10 s. In this case, the typical period of {\alf}ic pulses of current interest is {\sm}40 s \citep{OkaDePon2011}, and significant reflection of them is expected. 
We calculated the transmission coefficient using a classical model \citep{Zhugzhda1982}, which approximates the stratified solar atmosphere by a linear variation of temperature and an exponential variation of density. 
The transmission coefficient decreases with increasing {\alf} speed contrast, $\alpha= V_{A_2} /V_{A_1}$, across the transition layer, and also with decreasing density scale height, $H_\rho$.
Figure \ref{f7} shows the calculated transmission coefficients. In the left panel, the transmission reduces with increasing {\alf} speed contrast, $\alpha= V_{A_2} /V_{A_1}$, across the transition layer, and also with decreasing density scale height, $H_\rho$.
The right panel shows more systematically how the transmission coefficient depends on 
$\alpha$ and $H_\rho$ at a fixed period.
For a given $H_\rho$, lowering $\alpha$ increases the transmission coefficient.
At a given $\alpha$, a larger $H_\rho$ allows better transmission of waves, because it effectively increases $\omega/\omega_c$.  
For plausible ranges of $10\leq \alpha \leq 40$  and $200\leq H_\rho \leq$ 400 km, we can expect transmission of {\alf}ic pulses through the transition layer as high as 1--10\%, which is compatible with previous studies \citep{Schwartz1984, Cranmer2005, ChaeLee2023}.

\begin{thebibliography}{}
\expandafter\ifx\csname natexlab\endcsname\relax\def\natexlab#1{#1}\fi
\providecommand{\url}[1]{\href{#1}{#1}}
\providecommand{\dodoi}[1]{doi:~\href{http://doi.org/#1}{\nolinkurl{#1}}}
\providecommand{\doeprint}[1]{\href{http://ascl.net/#1}{\nolinkurl{http://ascl.net/#1}}}
\providecommand{\doarXiv}[1]{\href{https://arxiv.org/abs/#1}{\nolinkurl{https://arxiv.org/abs/#1}}}

\bibitem[{{Abramenko} {et~al.}(2013){Abramenko}, {Zank}, {Dosch}, {Yurchyshyn}, {Goode}, {Ahn}, \& {Cao}}]{Abramenko2013}
{Abramenko}, V.~I., {Zank}, G.~P., {Dosch}, A., {et~al.} 2013, \apj, 773, 167, \dodoi{10.1088/0004-637X/773/2/167}

\bibitem[{{Akhavan-Tafti} \& {Soni}(2024)}]{Akhavan2024}
{Akhavan-Tafti}, M., \& {Soni}, S.~L. 2024, \apjl, 970, L26, \dodoi{10.3847/2041-8213/ad60bc}

\bibitem[{{Asgari-Targhi} {et~al.}(2013){Asgari-Targhi}, {van Ballegooijen}, {Cranmer}, \& {DeLuca}}]{Asgari2013}
{Asgari-Targhi}, M., {van Ballegooijen}, A.~A., {Cranmer}, S.~R., \& {DeLuca}, E.~E. 2013, \apj, 773, 111, \dodoi{10.1088/0004-637X/773/2/111}

\bibitem[{{Bale} {et~al.}(2019){Bale}, {Badman}, {Bonnell}, {Bowen}, {Burgess}, {Case}, {Cattell}, {Chandran}, {Chaston}, {Chen}, {Drake}, {de Wit}, {Eastwood}, {Ergun}, {Farrell}, {Fong}, {Goetz}, {Goldstein}, {Goodrich}, {Harvey}, {Horbury}, {Howes}, {Kasper}, {Kellogg}, {Klimchuk}, {Korreck}, {Krasnoselskikh}, {Krucker}, {Laker}, {Larson}, {MacDowall}, {Maksimovic}, {Malaspina}, {Martinez-Oliveros}, {McComas}, {Meyer-Vernet}, {Moncuquet}, {Mozer}, {Phan}, {Pulupa}, {Raouafi}, {Salem}, {Stansby}, {Stevens}, {Szabo}, {Velli}, {Woolley}, \& {Wygant}}]{Bale2019}
{Bale}, S.~D., {Badman}, S.~T., {Bonnell}, J.~W., {et~al.} 2019, \nat, 576, 237, \dodoi{10.1038/s41586-019-1818-7}

\bibitem[{{Bale} {et~al.}(2021){Bale}, {Horbury}, {Velli}, {Desai}, {Halekas}, {McManus}, {Panasenco}, {Badman}, {Bowen}, {Chandran}, {Drake}, {Kasper}, {Laker}, {Mallet}, {Matteini}, {Phan}, {Raouafi}, {Squire}, {Woodham}, \& {Woolley}}]{Bale2021}
{Bale}, S.~D., {Horbury}, T.~S., {Velli}, M., {et~al.} 2021, \apj, 923, 174, \dodoi{10.3847/1538-4357/ac2d8c}

\bibitem[{{Bale} {et~al.}(2023){Bale}, {Drake}, {McManus}, {Desai}, {Badman}, {Larson}, {Swisdak}, {Horbury}, {Raouafi}, {Phan}, {Velli}, {McComas}, {Cohen}, {Mitchell}, {Panasenco}, \& {Kasper}}]{Bale2023}
{Bale}, S.~D., {Drake}, J.~F., {McManus}, M.~D., {et~al.} 2023, \nat, 618, 252, \dodoi{10.1038/s41586-023-05955-3}

\bibitem[{{Berberyan} {et~al.}(2024){Berberyan}, {Keys}, {Jess}, \& {Christian}}]{Berberyan2024}
{Berberyan}, A., {Keys}, P.~H., {Jess}, D.~B., \& {Christian}, D.~J. 2024, \aap, 690, A363, \dodoi{10.1051/0004-6361/202451199}

\bibitem[{{Bizien} {et~al.}(2025){Bizien}, {Froment}, {Madjarska}, {Dudok de Wit}, \& {Velli}}]{Bizien2025}
{Bizien}, N., {Froment}, C., {Madjarska}, M.~S., {Dudok de Wit}, T., \& {Velli}, M. 2025, \aap, 694, A181, \dodoi{10.1051/0004-6361/202452140}

\bibitem[{{Bowen} {et~al.}(2025){Bowen}, {Mallet}, {Dunn}, {Squire}, {Chandran}, {Meyrand}, {Davis}, {Dudok de Wit}, {Bale}, {Badman}, \& {Sioulas}}]{Bowen2025}
{Bowen}, T.~A., {Mallet}, A., {Dunn}, C.~I., {et~al.} 2025, arXiv e-prints, arXiv:2504.13384, \dodoi{10.48550/arXiv.2504.13384}

\bibitem[{Cao {et~al.}(2012)Cao, Goode, Ahn, Gorceix, Schmidt, \& Lin}]{Cao2012}
Cao, W., Goode, P.~R., Ahn, K., {et~al.} 2012, in Astronomical Society of the Pacific Conference Series, Vol. 463, Second ATST-EAST Meeting: Magnetic Fields from the Photosphere to the Corona., ed. T.~R. {Rimmele}, A.~{Tritschler}, F.~{W{\"o}ger}, M.~{Collados Vera}, H.~{Socas-Navarro}, R.~{Schlichenmaier}, M.~{Carlsson}, T.~{Berger}, A.~{Cadavid}, P.~R. {Gilbert}, P.~R. {Goode}, \& M.~{Kn{\"o}lker}, 291.
\newblock \url{https://ui.adsabs.harvard.edu/abs/2012ASPC..463..291C}

\bibitem[{Cao {et~al.}(2010)Cao, Gorceix, Coulter, Ahn, Rimmele, \& Goode}]{Cao2010}
Cao, W., Gorceix, N., Coulter, R., {et~al.} 2010, Astronomische Nachrichten, 331, 636, \dodoi{10.1002/asna.201011390}

\bibitem[{{Cattell} {et~al.}(2022){Cattell}, {Breneman}, {Dombeck}, {Hanson}, {Johnson}, {Halekas}, {Bale}, {Dudok de Wit}, {Goetz}, {Goodrich}, {Malaspina}, {Pulupa}, {Case}, {Kasper}, {Larson}, {Stevens}, \& {Whittlesey}}]{Cattell2022}
{Cattell}, C., {Breneman}, A., {Dombeck}, J., {et~al.} 2022, \apjl, 924, L33, \dodoi{10.3847/2041-8213/ac4015}

\bibitem[{{Chae} \& {Lee}(2023)}]{ChaeLee2023}
{Chae}, J., \& {Lee}, K.-S. 2023, \apj, 954, 45, \dodoi{10.3847/1538-4357/ace771}

\bibitem[{{Chitta} {et~al.}(2012){Chitta}, {van Ballegooijen}, {Rouppe van der Voort}, {DeLuca}, \& {Kariyappa}}]{Chitta2012}
{Chitta}, L.~P., {van Ballegooijen}, A.~A., {Rouppe van der Voort}, L., {DeLuca}, E.~E., \& {Kariyappa}, R. 2012, \apj, 752, 48, \dodoi{10.1088/0004-637X/752/1/48}

\bibitem[{{Choi} {et~al.}(2024){Choi}, {Agapitov}, {Colomban}, {Bonnell}, {Mozer}, {Sydora}, {Raouafi}, \& {Dudok de Wit}}]{Choi2024}
{Choi}, K.-E., {Agapitov}, O., {Colomban}, L., {et~al.} 2024, \apj, 971, 177, \dodoi{10.3847/1538-4357/ad54c4}

\bibitem[{{Choi} {et~al.}(2025){Choi}, {Agapitov}, {Lee}, {Mozer}, {Huang}, {Colomban}, {Verniero}, \& {Raouafi}}]{2025arXiv250608278C}
{Choi}, K.-E., {Agapitov}, O.~V., {Lee}, D.-Y., {et~al.} 2025, \apj, arXiv:2506.08278, \dodoi{10.48550/arXiv.2506.08278}

\bibitem[{{Cranmer} \& {van Ballegooijen}(2005)}]{Cranmer2005}
{Cranmer}, S.~R., \& {van Ballegooijen}, A.~A. 2005, \apjs, 156, 265, \dodoi{10.1086/426507}

\bibitem[{{Cranmer} \& {Winebarger}(2019)}]{Cranmer2019}
{Cranmer}, S.~R., \& {Winebarger}, A.~R. 2019, \araa, 57, 157, \dodoi{10.1146/annurev-astro-091918-104416}

\bibitem[{{DeForest} {et~al.}(2007){DeForest}, {Hagenaar}, {Lamb}, {Parnell}, \& {Welsch}}]{DeForest2007}
{DeForest}, C.~E., {Hagenaar}, H.~J., {Lamb}, D.~A., {Parnell}, C.~E., \& {Welsch}, B.~T. 2007, \apj, 666, 576, \dodoi{10.1086/518994}

\bibitem[{{Dudok de Wit} {et~al.}(2020){Dudok de Wit}, {Krasnoselskikh}, {Bale}, {Bonnell}, {Bowen}, {Chen}, {Froment}, {Goetz}, {Harvey}, {Jagarlamudi}, {Larosa}, {MacDowall}, {Malaspina}, {Matthaeus}, {Pulupa}, {Velli}, \& {Whittlesey}}]{deWit2020}
{Dudok de Wit}, T., {Krasnoselskikh}, V.~V., {Bale}, S.~D., {et~al.} 2020, \apjs, 246, 39, \dodoi{10.3847/1538-4365/ab5853}

\bibitem[{{Dunn} \& {Zirker}(1973)}]{DunnZir1973}
{Dunn}, R.~B., \& {Zirker}, J.~B. 1973, \solphys, 33, 281, \dodoi{10.1007/BF00152419}

\bibitem[{{Fargette} {et~al.}(2021){Fargette}, {Lavraud}, {Rouillard}, {R{\'e}ville}, {Dudok De Wit}, {Froment}, {Halekas}, {Phan}, {Malaspina}, {Bale}, {Kasper}, {Louarn}, {Case}, {Korreck}, {Larson}, {Pulupa}, {Stevens}, {Whittlesey}, \& {Berthomier}}]{Fargette2021}
{Fargette}, N., {Lavraud}, B., {Rouillard}, A.~P., {et~al.} 2021, \apj, 919, 96, \dodoi{10.3847/1538-4357/ac1112}

\bibitem[{{Freeland} \& {Handy}(2012)}]{2012ascl.soft08013F}
{Freeland}, S.~L., \& {Handy}, B.~N. 2012, {SolarSoft: Programming and data analysis environment for solar physics}, Astrophysics Source Code Library, record ascl:1208.013

\bibitem[{{Georgoulis} {et~al.}(2025){Georgoulis}, {Li}, {Lee}, {Wang}, \& {Raouafi}}]{Manolis25}
{Georgoulis}, M.~K., {Li}, Q., {Lee}, J., {Wang}, H., \& {Raouafi}, N. 2025, \apj, in preparation

\bibitem[{{Georgoulis} {et~al.}(2012){Georgoulis}, {Titov}, \& {Miki{\'c}}}]{2012Georgoulis}
{Georgoulis}, M.~K., {Titov}, V.~S., \& {Miki{\'c}}, Z. 2012, \apj, 761, 61, \dodoi{10.1088/0004-637X/761/1/61}

\bibitem[{Goode \& Cao(2012)}]{Goode2012}
Goode, P.~R., \& Cao, W. 2012, in Society of Photo-Optical Instrumentation Engineers (SPIE) Conference Series, Vol. 8444, Ground-based and Airborne Telescopes IV, ed. L.~M. {Stepp}, R.~{Gilmozzi}, \& H.~J. {Hall}, 844403, \dodoi{10.1117/12.925494}

\bibitem[{{Hollweg}(1978)}]{Hollweg1978}
{Hollweg}, J.~V. 1978, \solphys, 56, 305, \dodoi{10.1007/BF00152474}

\bibitem[{{Howard} {et~al.}(2023){Howard}, {Vourlidas}, \& {Stenborg}}]{2023Howard}
{Howard}, R.~A., {Vourlidas}, A., \& {Stenborg}, G. 2023, Frontiers in Astronomy and Space Sciences, 10, 1264226, \dodoi{10.3389/fspas.2023.1264226}

\bibitem[{{Jess} {et~al.}(2012){Jess}, {Pascoe}, {Christian}, {Mathioudakis}, {Keys}, \& {Keenan}}]{Jess2012}
{Jess}, D.~B., {Pascoe}, D.~J., {Christian}, D.~J., {et~al.} 2012, \apjl, 744, L5, \dodoi{10.1088/2041-8205/744/1/L5}

\bibitem[{{Johnston} {et~al.}(2022){Johnston}, {Squire}, {Mallet}, \& {Meyrand}}]{Johnston2022}
{Johnston}, Z., {Squire}, J., {Mallet}, A., \& {Meyrand}, R. 2022, Physics of Plasmas, 29, 072902, \dodoi{10.1063/5.0097983}

\bibitem[{{Kalkofen}(1997)}]{Kalkofen1997}
{Kalkofen}, W. 1997, \apjl, 486, L145, \dodoi{10.1086/310842}

\bibitem[{{Kasper} {et~al.}(2019){Kasper}, {Bale}, {Belcher}, {Berthomier}, {Case}, {Chandran}, {Curtis}, {Gallagher}, {Gary}, {Golub}, {Halekas}, {Ho}, {Horbury}, {Hu}, {Huang}, {Klein}, {Korreck}, {Larson}, {Livi}, {Maruca}, {Lavraud}, {Louarn}, {Maksimovic}, {Martinovic}, {McGinnis}, {Pogorelov}, {Richardson}, {Skoug}, {Steinberg}, {Stevens}, {Szabo}, {Velli}, {Whittlesey}, {Wright}, {Zank}, {MacDowall}, {McComas}, {McNutt}, {Pulupa}, {Raouafi}, \& {Schwadron}}]{Kasper2019}
{Kasper}, J.~C., {Bale}, S.~D., {Belcher}, J.~W., {et~al.} 2019, \nat, 576, 228, \dodoi{10.1038/s41586-019-1813-z}

\bibitem[{{Krasnoselskikh} {et~al.}(2020){Krasnoselskikh}, {Larosa}, {Agapitov}, {de Wit}, {Moncuquet}, {Mozer}, {Stevens}, {Bale}, {Bonnell}, {Froment}, {Goetz}, {Goodrich}, {Harvey}, {Kasper}, {MacDowall}, {Malaspina}, {Pulupa}, {Raouafi}, {Revillet}, {Velli}, \& {Wygant}}]{Krasnoselskikh2020}
{Krasnoselskikh}, V., {Larosa}, A., {Agapitov}, O., {et~al.} 2020, \apj, 893, 93, \dodoi{10.3847/1538-4357/ab7f2d}

\bibitem[{{Laker} {et~al.}(2021){Laker}, {Horbury}, {Bale}, {Matteini}, {Woolley}, {Woodham}, {Badman}, {Pulupa}, {Kasper}, {Stevens}, {Case}, \& {Korreck}}]{Laker2021}
{Laker}, R., {Horbury}, T.~S., {Bale}, S.~D., {et~al.} 2021, \aap, 650, A1, \dodoi{10.1051/0004-6361/202039354}

\bibitem[{{Larosa} {et~al.}(2024){Larosa}, {Chen}, {McIntyre}, {Jagarlamudi}, \& {Sorriso-Valvo}}]{Larosa2024}
{Larosa}, A., {Chen}, C.~H.~K., {McIntyre}, J.~R., {Jagarlamudi}, V.~K., \& {Sorriso-Valvo}, L. 2024, \aap, 686, A238, \dodoi{10.1051/0004-6361/202450030}

\bibitem[{{Larosa} {et~al.}(2021){Larosa}, {Krasnoselskikh}, {Dudok de Wit}, {Agapitov}, {Froment}, {Jagarlamudi}, {Velli}, {Bale}, {Case}, {Goetz}, {Harvey}, {Kasper}, {Korreck}, {Larson}, {MacDowall}, {Malaspina}, {Pulupa}, {Revillet}, \& {Stevens}}]{Larosa2021}
{Larosa}, A., {Krasnoselskikh}, V., {Dudok de Wit}, T., {et~al.} 2021, \aap, 650, A3, \dodoi{10.1051/0004-6361/202039442}

\bibitem[{{Lee}(2022)}]{2022LeeRvMPP}
{Lee}, J. 2022, Reviews of Modern Plasma Physics, 6, 32, \dodoi{10.1007/s41614-022-00096-y}

\bibitem[{{Lee} {et~al.}(2024){Lee}, {Wang}, {Wang}, \& {Wang}}]{Leej2024}
{Lee}, J., {Wang}, H., {Wang}, J., \& {Wang}, M. 2024, \apj, 963, 79, \dodoi{10.3847/1538-4357/ad23e0}

\bibitem[{{Lee} {et~al.}(2022){Lee}, {Yurchyshyn}, {Wang}, {Yang}, {Cao}, \& {Carlos Mart{\'\i}nez Oliveros}}]{Leej2022}
{Lee}, J., {Yurchyshyn}, V., {Wang}, H., {et~al.} 2022, \apjl, 935, L27, \dodoi{10.3847/2041-8213/ac86bf}

\bibitem[{{Mackenzie Dover} {et~al.}(2021){Mackenzie Dover}, {Sharma}, \& {Erd{\'e}lyi}}]{Dover2021}
{Mackenzie Dover}, F., {Sharma}, R., \& {Erd{\'e}lyi}, R. 2021, \apj, 913, 19, \dodoi{10.3847/1538-4357/abefd1}

\bibitem[{{Mackenzie Dover} {et~al.}(2022){Mackenzie Dover}, {Sharma}, \& {Erd{\'e}lyi}}]{Dover2022}
---. 2022, \apj, 929, 88, \dodoi{10.3847/1538-4357/ac5aa9}

\bibitem[{{Mallet} {et~al.}(2021){Mallet}, {Squire}, {Chandran}, {Bowen}, \& {Bale}}]{Mallet2021}
{Mallet}, A., {Squire}, J., {Chandran}, B. D.~G., {Bowen}, T., \& {Bale}, S.~D. 2021, \apj, 918, 62, \dodoi{10.3847/1538-4357/ac0c12}

\bibitem[{{Nisenson} {et~al.}(2003){Nisenson}, {van Ballegooijen}, {de Wijn}, \& {S{\"u}tterlin}}]{Nisenson2003}
{Nisenson}, P., {van Ballegooijen}, A.~A., {de Wijn}, A.~G., \& {S{\"u}tterlin}, P. 2003, \apj, 587, 458, \dodoi{10.1086/368067}

\bibitem[{{Okamoto} \& {De Pontieu}(2011)}]{OkaDePon2011}
{Okamoto}, T.~J., \& {De Pontieu}, B. 2011, \apjl, 736, L24, \dodoi{10.1088/2041-8205/736/2/L24}

\bibitem[{{Parker}(1958)}]{Parker1958}
{Parker}, E.~N. 1958, \apj, 128, 664, \dodoi{10.1086/146579}

\bibitem[{{Parker}(1965)}]{Parker1965}
---. 1965, \ssr, 4, 666, \dodoi{10.1007/BF00216273}

\bibitem[{{Parnell}(2002)}]{Parnell2002}
{Parnell}, C.~E. 2002, \mnras, 335, 389, \dodoi{10.1046/j.1365-8711.2002.05618.x}

\bibitem[{{Parnell} {et~al.}(2009){Parnell}, {DeForest}, {Hagenaar}, {Johnston}, {Lamb}, \& {Welsch}}]{Parnell2009}
{Parnell}, C.~E., {DeForest}, C.~E., {Hagenaar}, H.~J., {et~al.} 2009, \apj, 698, 75, \dodoi{10.1088/0004-637X/698/1/75}

\bibitem[{{Raouafi} {et~al.}(2023{\natexlab{a}}){Raouafi}, {Stenborg}, {Seaton}, {Wang}, {Wang}, {DeForest}, {Bale}, {Drake}, {Uritsky}, {Karpen}, {DeVore}, {Sterling}, {Horbury}, {Harra}, {Bourouaine}, {Kasper}, {Kumar}, {Phan}, \& {Velli}}]{Raouafi2023}
{Raouafi}, N.~E., {Stenborg}, G., {Seaton}, D.~B., {et~al.} 2023{\natexlab{a}}, \apj, 945, 28, \dodoi{10.3847/1538-4357/acaf6c}

\bibitem[{{Raouafi} {et~al.}(2023{\natexlab{b}}){Raouafi}, {Matteini}, {Squire}, {Badman}, {Velli}, {Klein}, {Chen}, {Matthaeus}, {Szabo}, {Linton}, {Allen}, {Szalay}, {Bruno}, {Decker}, {Akhavan-Tafti}, {Agapitov}, {Bale}, {Bandyopadhyay}, {Battams}, {Ber{\v{c}}i{\v{c}}}, {Bourouaine}, {Bowen}, {Cattell}, {Chandran}, {Chhiber}, {Cohen}, {D'Amicis}, {Giacalone}, {Hess}, {Howard}, {Horbury}, {Jagarlamudi}, {Joyce}, {Kasper}, {Kinnison}, {Laker}, {Liewer}, {Malaspina}, {Mann}, {McComas}, {Niembro-Hernandez}, {Nieves-Chinchilla}, {Panasenco}, {Pokorn{\'y}}, {Pusack}, {Pulupa}, {Perez}, {Riley}, {Rouillard}, {Shi}, {Stenborg}, {Tenerani}, {Verniero}, {Viall}, {Vourlidas}, {Wood}, {Woodham}, \& {Woolley}}]{Raouafi2023SSRv}
{Raouafi}, N.~E., {Matteini}, L., {Squire}, J., {et~al.} 2023{\natexlab{b}}, \ssr, 219, 8, \dodoi{10.1007/s11214-023-00952-4}

\bibitem[{{Rivera} {et~al.}(2024){Rivera}, {Badman}, {Stevens}, {Verniero}, {Stawarz}, {Shi}, {Raines}, {Paulson}, {Owen}, {Niembro}, {Louarn}, {Livi}, {Lepri}, {Kasper}, {Horbury}, {Halekas}, {Dewey}, {De Marco}, \& {Bale}}]{Rivera2024}
{Rivera}, Y.~J., {Badman}, S.~T., {Stevens}, M.~L., {et~al.} 2024, Science, 385, 962, \dodoi{10.1126/science.adk6953}

\bibitem[{{Schwartz} \& {Bel}(1984)}]{Schwartz1984}
{Schwartz}, S.~J., \& {Bel}, N. 1984, \aap, 137, 128

\bibitem[{{Secchi}(1877)}]{Secchi1877}
{Secchi}, A. 1877, {L'astronomia in Roma nel pontificato DI Pio IX.}

\bibitem[{{Sharma} \& {Morton}(2023)}]{Sharma2023}
{Sharma}, R., \& {Morton}, R.~J. 2023, Nature Astronomy, 7, 1301, \dodoi{10.1038/s41550-023-02070-1}

\bibitem[{{Shi} {et~al.}(2022){Shi}, {Panasenco}, {Velli}, {Tenerani}, {Verniero}, {Sioulas}, {Huang}, {Brosius}, {Bale}, {Klein}, {Kasper}, {de Wit}, {Goetz}, {Harvey}, {MacDowall}, {Malaspina}, {Pulupa}, {Larson}, {Livi}, {Case}, \& {Stevens}}]{Shi2022}
{Shi}, C., {Panasenco}, O., {Velli}, M., {et~al.} 2022, \apj, 934, 152, \dodoi{10.3847/1538-4357/ac7c11}

\bibitem[{{Shoda} {et~al.}(2021){Shoda}, {Chandran}, \& {Cranmer}}]{Shoda2021}
{Shoda}, M., {Chandran}, B. D.~G., \& {Cranmer}, S.~R. 2021, \apj, 915, 52, \dodoi{10.3847/1538-4357/abfdbc}

\bibitem[{{Simon} \& {Leighton}(1964)}]{Simon1964}
{Simon}, G.~W., \& {Leighton}, R.~B. 1964, \apj, 140, 1120, \dodoi{10.1086/148010}

\bibitem[{{Soler} {et~al.}(2019){Soler}, {Terradas}, {Oliver}, \& {Ballester}}]{Soler2019}
{Soler}, R., {Terradas}, J., {Oliver}, R., \& {Ballester}, J.~L. 2019, \apj, 871, 3, \dodoi{10.3847/1538-4357/aaf64c}

\bibitem[{{Squire} {et~al.}(2020){Squire}, {Chandran}, \& {Meyrand}}]{Squire2020}
{Squire}, J., {Chandran}, B.~D.~G., \& {Meyrand}, R. 2020, \apjl, 891, L2, \dodoi{10.3847/2041-8213/ab74e1}

\bibitem[{{Squire} {et~al.}(2022){Squire}, {Johnston}, {Mallet}, \& {Meyrand}}]{Squire2022}
{Squire}, J., {Johnston}, Z., {Mallet}, A., \& {Meyrand}, R. 2022, Physics of Plasmas, 29, 112903, \dodoi{10.1063/5.0099924}

\bibitem[{{Stix}(2004)}]{Stix2004}
{Stix}, M. 2004, {The Sun: An Introduction}

\bibitem[{{Touresse} {et~al.}(2024){Touresse}, {Pariat}, {Froment}, {Aslanyan}, {Wyper}, \& {Seyfritz}}]{Touresse2024}
{Touresse}, J., {Pariat}, E., {Froment}, C., {et~al.} 2024, \aap, 692, A71, \dodoi{10.1051/0004-6361/202452019}

\bibitem[{{Ulmschneider} {et~al.}(1991){Ulmschneider}, {Zaehringer}, \& {Musielak}}]{Ulmschneider1991}
{Ulmschneider}, P., {Zaehringer}, K., \& {Musielak}, Z.~E. 1991, \aap, 241, 625

\bibitem[{{Utz} {et~al.}(2010){Utz}, {Hanslmeier}, {Muller}, {Veronig}, {Ryb{\'a}k}, \& {Muthsam}}]{Utz2010}
{Utz}, D., {Hanslmeier}, A., {Muller}, R., {et~al.} 2010, \aap, 511, A39, \dodoi{10.1051/0004-6361/200913085}

\bibitem[{{van Ballegooijen} {et~al.}(1998){van Ballegooijen}, {Nisenson}, {Noyes}, {L{\"o}fdahl}, {Stein}, {Nordlund}, \& {Krishnakumar}}]{Ballegoo1998}
{van Ballegooijen}, A.~A., {Nisenson}, P., {Noyes}, R.~W., {et~al.} 1998, \apj, 509, 435, \dodoi{10.1086/306471}

\bibitem[{{Verscharen} {et~al.}(2019){Verscharen}, {Klein}, \& {Maruca}}]{Verscharen2019}
{Verscharen}, D., {Klein}, K.~G., \& {Maruca}, B.~A. 2019, Living Reviews in Solar Physics, 16, 5, \dodoi{10.1007/s41116-019-0021-0}

\bibitem[{{Viall} {et~al.}(2021){Viall}, {DeForest}, \& {Kepko}}]{Viall2021}
{Viall}, N.~M., {DeForest}, C.~E., \& {Kepko}, L. 2021, Frontiers in Astronomy and Space Sciences, 8, 139, \dodoi{10.3389/fspas.2021.735034}

\bibitem[{{Wang} \& {Sheeley}(1990)}]{Wangs1990}
{Wang}, Y.~M., \& {Sheeley}, N.~R., J. 1990, \apj, 355, 726, \dodoi{10.1086/168805}

\bibitem[{{Wedemeyer-B{\"o}hm} {et~al.}(2012){Wedemeyer-B{\"o}hm}, {Scullion}, {Steiner}, {Rouppe van der Voort}, {de La Cruz Rodriguez}, {Fedun}, \& {Erd{\'e}lyi}}]{Wedemeyer2012}
{Wedemeyer-B{\"o}hm}, S., {Scullion}, E., {Steiner}, O., {et~al.} 2012, \nat, 486, 505, \dodoi{10.1038/nature11202}

\bibitem[{{Wyper} {et~al.}(2024){Wyper}, {Pariat}, {Squire}, {Touresse}, \& {Matteini}}]{Wyper2024AGU01W}
{Wyper}, P.~F., {Pariat}, E., {Squire}, J., {Touresse}, J., \& {Matteini}, L. 2024, in AGU Fall Meeting Abstracts, Vol. 2024, SH24A--01

\bibitem[{{Zhugzhda} \& {Locans}(1982)}]{Zhugzhda1982}
{Zhugzhda}, I.~D., \& {Locans}, V. 1982, \solphys, 76, 77, \dodoi{10.1007/BF00214131}

\bibitem[{{Zirin}(1988)}]{Zirin1988}
{Zirin}, H. 1988, {Astrophysics of the sun}

\end{thebibliography}
\end{document}